\documentclass[aps,prd,a4paper,twocolumn,floatfix,nofootinbib]{revtex4}
\usepackage{amssymb}  
\usepackage{floatflt}
\usepackage{graphicx}%[dvips,dvipdfm,pdftex]
\usepackage{epsfig}
\usepackage{natbib}
\usepackage{color}
\DeclareGraphicsExtensions{.pdf,.png,.jpg}

\begin{document}

\def\pp{{\, \mid \hskip -1.5mm =}} 
\def\cL{{\cal L}}  
\def\beq{\begin{equation}} 
\def\eneq{\end{equation}} 
\def\bea{\begin{eqnarray}} 
\def\enea{\end{eqnarray}} 
\def\tr{{\rm tr}\, } 
\def\nn{\nonumber \\} 
\def\e{{\rm e}}

\title{\textbf{The density profiles of Dark Matter halos in Spiral Galaxies}} 
 
\author{Gianluca Castignani, Noemi Frusciante, Daniele Vernieri, Paolo Salucci} 

\affiliation{SISSA--ISAS, International School for Advanced Studies, Via Bonomea 265, 34136, Trieste, Italy {\rm and} INFN, Sezione di Trieste, Via Valerio 2, 34127, Trieste, Italy}

\maketitle

\subparagraph{ABSTRACT \\ \\} 
{\bf In spiral galaxies,  we  explain their non-Keplerian rotation curves (RCs) by means of  a non-luminous component embedding their stellar-gaseous disks.  Understanding  the  detailed  properties of this  component   (labelled Dark Matter, DM) is one of the most pressing issues of Cosmology.
We investigate the recent  relationship (claimed by Walker et al. 2010, hereafter W+10) between   $r$, the galaxy  radial coordinate,  and $V_h(r)$, the dark  halo contribution to the circular velocity at $r$, {\it  a})  in the framework of the Universal Rotation Curve (URC) paradigm  and  directly {\it  b})  by means  of the kinematics of a large  sample  of  DM   dominated spirals.
We find  a general  agreement between the W+10 claim, the distribution of DM  emerging from the  URC and that inferred in the (low luminosity)  objects of our sample. We show that such a phenomenology, linking the spiral's  luminosity, radii and circular velocities,  implies  an evident  inconsistency with (naive) predictions in the $\Lambda$ Cold Dark Matter ($\Lambda$CDM) scenario.}   

\section{Introduction}

\vspace{0.4cm}

Rotation curves (RCs) of spiral galaxies show no asymptotic  Keplerian behavior and fail to match  the distribution of the luminous matter. The favored   explanation is the existence of an unknown unseen component, the Dark Matter (DM): the  ``luminous'' components of galaxies such as gas and stars are
 distributed  inside  a spherical ``dark'' halo \footnote{A thorough review on this issue can be found in \cite{dmaw} and downloaded at www.sissa.it/ap/dmg/dmaw\_presentation.html.}. \\
Recently,  \cite{walker}, hereafter  W+10,  following the pioneer paper of  \cite{mcg2007}  (M+07), found that the dark component of the circular velocity  (and of  mass) profile  shows a characteristic and  similar  behavior both in Spirals (and in Milky Way~dSphs) of very different luminosities.  More precisely, they  investigated a sample of 60 spirals, whose RCs span  radii from 1 kpc to  75 kpc,  have  maximum amplitudes   in the range   $50  \mbox{ km s}^{-1} \leqslant  V_{max} \leqslant 300 \mbox{ km s}^{-1}$ and  their  baryonic masses  $M_b $ ranging   between  3 $\times 10^8 $ M$_{\odot} $ and $4\times 10^{11}$ M$_{\odot}$.   They derived $V_h(r)$, the halo  contributions to the circular velocity (in short the ``halo RCs''),   by subtracting the contribution of the baryonic matter from the circular velocity: $V^2_h(r)= V^2 (r) - V^2_{b}(r)$ (see W+10 for details).   They  found  a correlation between  $V_h(r)$ and $r$  (see  W+10):    
\beq
\log \left( \frac{V_h(r) }{\mbox{km\,s}^{-1}}\right) =1.47^{+0.15}_{-0.19}+0.5\log \left( \frac{r}{\mbox{kpc}}\right),          
       \label{eq:0.2}
\eneq
{\it independently}  of galaxy luminosity. Moreover:  $\frac{M_h(r)}{M_{\odot}}=2^{+2}_{-1.2} \times 10^8\left( \frac{r}{kpc}\right)^2.$   Noticeably,  at $r=R_{1/2}$, being $R_{1/2}$  the radius encompassing half of the {\it stellar} mass, the relation (\ref{eq:0.2}) can be investigated also in dwarf spheroidals, and  found to hold also in these systems (see W+10 for details). \\
These  are  surprising findings:  in fact, when,   in objects of {\it very different}  luminosity we  derive their  $V_h(r)$ (the  {\it dark halo-contribution}   to the circular velocities) and  plot them  as a function of radius, a well defined curve emerges, instead of  a more likely scatter plot.  On the other hand,  the {\it circular velocities } $V(r)$ show  a very  different phenomenology. From the analysis of  several thousands of  RCs, an  {\it Universal Rotation Curve} (URC) emerges,  i.e.   a specific 3-dimensional surface linking  a) the circular velocity at a {\it normalized} radius $r/R_D$  (i.e. measured in units of  disk length-scale),  b) the galaxy  luminosity and c), the  normalized   radius   (see  \cite{persal91}, \cite{persalstel} (PSS), \cite{cat}, \cite{Rhee} and \cite{sallapi2007} (S+07) for details). \\
Summarizing, in the intermediate regions of spirals,   W+10 found:  $V_h(r)= F_{W+10}(r)$,  independently  of galaxy  luminosity $L$ and stellar disk length-scale $R_D$. In  the URC scenario, instead,  we have:   $V_h(r)=F_{URC}(r/R_D, L)$, i.e. the  halo component of the circular  velocity is  a function of disk luminosity and disk length scale.  Let us stress that the existence of the  URC implies that the structural parameters  of the  Dark  and  Luminous components are strongly related, but not necessarily, through  the very  constraining eq. (\ref{eq:0.2}). 

In this paper,  we investigate whether these  two apparently different results   are compatible. We  also use the outcome  to  derive   important  information on  the DM distribution  in  spirals.  We   will study an independent (larger) sample of spirals by means  of  reliable method of mass modelling.   From \cite{persal95} (PS95) we select $116$ spiral galaxies (hereafter Sample A) whose  RCs satisfy  the following requirements: a) no presence of a prominent bulge:  $V(0.2 R_{opt}) <  50\mbox{ km s}^{-1}$ \footnote{$R_{opt}$ is the radius encircling 83\% of the light, for a Freeman exponential disk $R_{opt} = 3.2R_D$ where $R_D$ is the disk  scale-length, see \cite{free}.}, or  if the latter is missing:   $V (0.4 R_{opt}) < 70 \mbox{ km s}^{-1}$, b) low luminosity objects:  $V(r)< 120 \mbox{ km s}^{-1}$, at any radius, that implies:  $M_I>-20.2$, with $M_I$ the magnitude in the I band, c) the RCs have a measure either at 0.8 $R_{opt}$ or  at 1.0 $R_{opt}$. \\
Let us notice that  we limit ourselves to   objects of low luminosity,  in that they are DM dominated (e.g. \cite{persal90} (PS90)) and therefore  $V_h(r)$ can be derived with great accuracy.  Higher luminosity spirals are disk dominated and therefore  $V_h(r)$ strongly depends on the details and on the assumptions in  the mass modelling.
The RCs  of Sample A  have a radial range matching that of  W+10 sample $\max\left[ \frac{1}{3}R_D, 1 \mbox{kpc}\right]\leqslant \,r \, \leqslant 6 R_D$. Following W+10,  we adopted  1 kpc as the  innermost  radius in that, inside which, the stellar disk (sometimes    with  a stellar  bulge)   dominates the gravitational potential  making  difficult the  estimate of  $V_h(r)$. \\
Let us  compare the  W+10 relationship   with the  URC analogue (PSS, S+07) and with that  we  derive  for  Sample A.   From  the condition of centrifugal equilibrium  we have:   
\beq
V^2(r)=V^2_{lum}(r)+V^2_{h}(r),                                  
\eneq
where the subscripts \textit{lum} and \textit{h} stand respectively for luminous and halo components.
The former is the quadratic sum of disk, bulge and gas contributions. Here,  we neglect the latter two because  1) the selected low luminosity objects  have,  for $r>1\mbox{ kpc}$, a negligible  bulge, 2) the gaseous disk is important only for $r> R_{opt}$, i.e.  outside the region that we are considering here (see \cite{evoli}).
For the objects of Sample A, each RC has 2-4 independent measurements (see Table 2 of PS95) and therefore the  same   number of halo velocity data.  For the URC,  there is analytical formula for  $V_h(r)$ (see below). 
Hereafter,  we adopt a flat cosmology with matter density parameter $\Omega_M=0.27$ and Hubble constant $H_0=75 \mbox{ km s}^{-1} \mbox{ Mpc}^{-1}$.

\section{The contribution of Dark Matter Halos to the circular velocity of Spirals} 

\begin{figure*}[htbp]
\includegraphics[width=14cm,height=16cm]{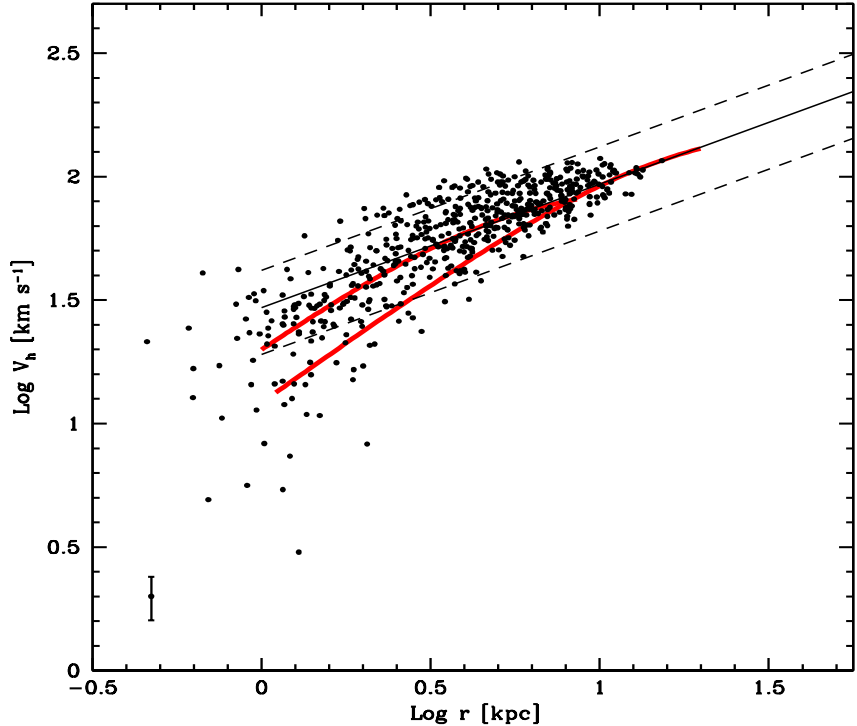}
\vskip -3.5cm
\caption{The W+10 halo velocity-radius  relationship (black line)  with its 1-$\sigma$ scatter (black dashed lines) plotted with 1)  the halo velocities of  our Sample A of 116 spirals (black points) (PSS, PS90),  2) the URCH  profiles (red lines),  corresponding to objects  with  mass comparable with those in the W+10 sample. The error (of 20 \%) in the individual determination  of $V_h$ is shown as an errorbar.}
\label{fig:2}
\end{figure*}

\begin{figure*}[htbp]
\includegraphics[width=14cm,height=16cm]{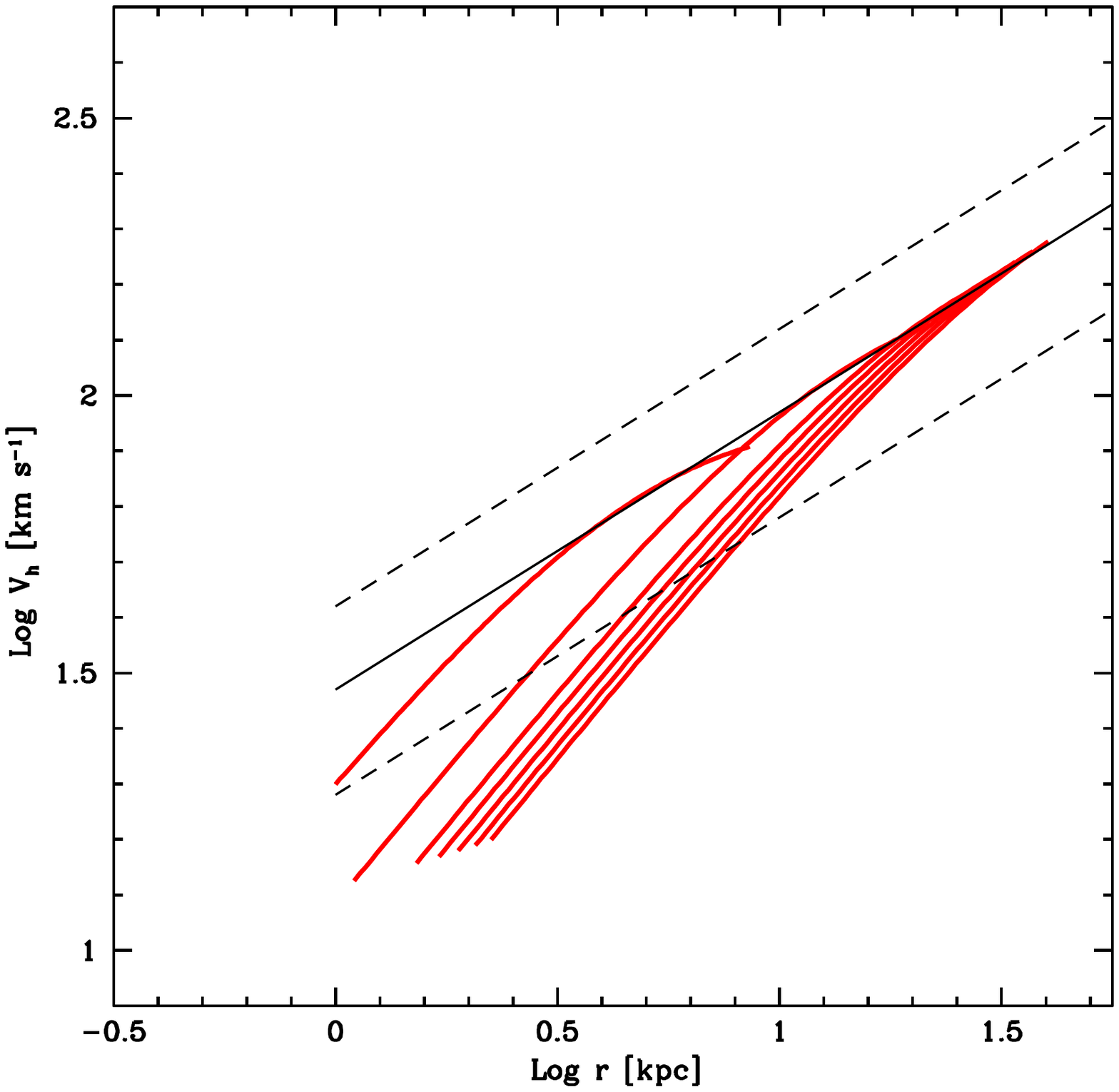}
\vskip -3.5cm
\caption{The W+10 halo velocity-radius  relationship (black line)  with its 1-$\sigma$ scatter (black dashed lines) plotted with the URCH  profiles (red lines).}
\label{fig:4}
\end{figure*}

We derive the velocity  halo contribution  for the objects of Sample A in the following way; we assume  a  Freeman profile to  describe  the stellar disk surface density, then  \cite{free}:
\beq
V^2_D(x)=14.9\, \beta(x) \, V^2 (1.0)\, x^2\,[I_0K_0-I_1K_1]_{1.6x},   \label{eq:0.1}
\eneq
where $x=r/R_{opt}$,  $I_n$ and $K_n$ are the modified Bessel functions. Let us also notice that:   $R_{1/2}= 1.67 /3.2 \, R_{opt}=0.52\, R_{opt}.$ 
The mass modelling of  individual (and coadded)  RCs   of objects of low luminosity  (as those in  Sample A) is rather simple.   We have: $\beta(1) \simeq 0.13$ and  $\beta(0.8) \simeq  0.23$ (see PS90 and PSS). More precisely,  there is a small luminosity dependence of $\beta$  but for the objects in  Sample A  this is irrelevant. Then, for any object,   we derive $V_h(r)$  by subtracting the  (small) disk  contribution  $V^2_D(r)$,  given by eq. (\ref{eq:0.1}),   from the  circular velocity  $V^2(r)$.    Finally, we derive   $V_h(R_{1/2})$ by linearly interpolating  $V_h(0.4 R_{opt})$ and  $V_h(0.6 R_{opt})$.  \\
The URC leads to  an (analytical) Universal  form for $V_h(r)$,  i.e. for the   {\it halo} velocity component  (URCH) to $V_{URC}(r)$ which was  built by  coadding the kinematics of thousands of galaxies (PSS, S+07). We  model the   URC,   that represents the typical RC  of an object of magnitude $M_I$, (or of  virial halo mass $M_{vir}$ (see S+07 and reference therein)  in its dark and luminous components (e.g. S+07) \footnote{The virial mass $M_{vir}$ corresponds to the mass enclosed in a sphere with density $101\,\rho_c$, being $\rho_c$ the critical density of the Universe. $M_{vir}$ and the virial radius are related by  $R_{vir}=259\left(M_{vir}/10^{12}M_\odot\right)^{1/3}$ kpc, see \cite{eke}.}).
\begin{figure*}[htbp]
\includegraphics[width=14cm,height=16cm]{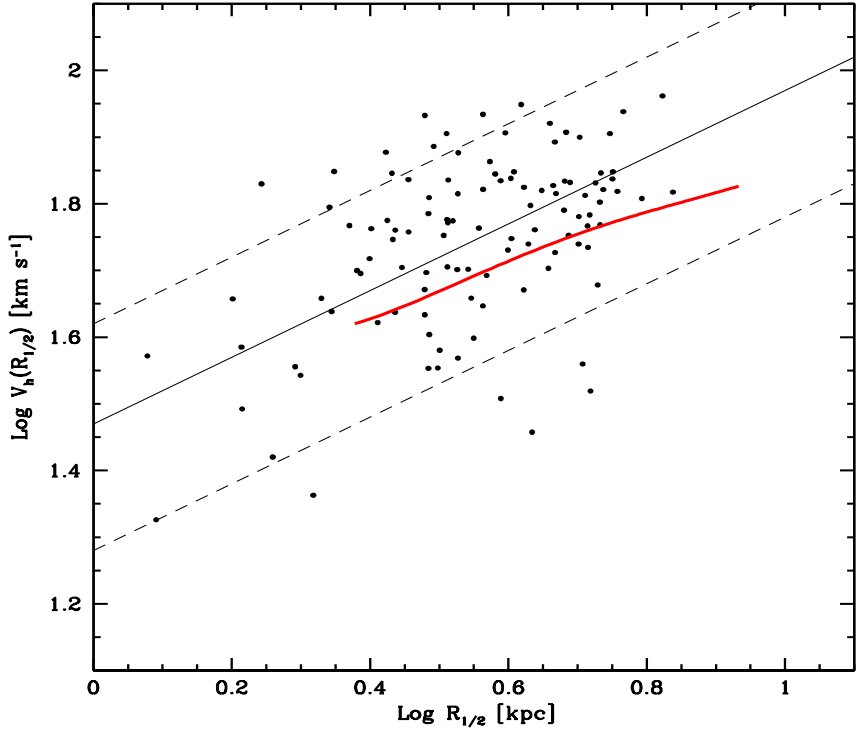}
\vskip -3.5cm
\caption{The W+10 halo velocity-radius relationship (black line) with its 1-$\sigma$ scatter (black dashed lines) compared with the halo velocity calculated at $R_{1/2}$  as function of $R_{1/2}$  for 1) our Sample A whose points are calculated by interpolating  $V_h(0.4 R_{opt})$ and  $V_h(0.6 R_{opt})$ for each galaxy (black points) and 2) URCH profiles (red  line).} 
\label{fig:1}
\end{figure*}
In detail the  URC  DM density profile is 
\beq
\rho_{URCH}(r)=\frac{\rho_0 r_0^3}{(r+r_0)(r^2+r_0^2)},  \label{eq:0.3}
\eneq     
where $r_0$ is the core radius and $\rho_0$ is the central halo density. It follows that: 
\bea	
V^2_{URCH}(r)&=&6.4\frac{\rho_0r_0^3}{r}\left[\ln\left(1+\frac{r}{r_0}\right)-\arctan\left(\frac{r}{r_0}\right)\right.\nonumber \\ 
&+&\left.\frac{1}{2}\ln\left(1+\frac{r^2}{r_0^2}\right)\right].                       
\enea 
From \cite{donato} we have:
\beq
\rho_0=5\times 10^{-24}(r_0/(8.6 \,\mbox{kpc}))^{-1} \mbox{g cm}^{-3},     \label{eq:0.4}         
\eneq
and 
\beq
\log \left( \frac{r_0}{\mbox{kpc}}\right) \simeq 0.66 +0.58 \log\left( \frac{M_{vir}}{10^{11}M_{\odot}}\right).                     
\eneq 
The disk length-scales and  masses,  $R_D$ and  $M_D$, are related to the halo masses through (see \cite{shank} and  S+07) 
\bea
\log \left( \frac{R_D}{\mbox{kpc}}\right) &=&0.633+0.379\log\left( \frac{M_D}{10^{11}M_{\odot}}\right)  \nonumber \\
&+&0.069\left(\log \frac{M_D}{10^{11}M_{\odot}} \right)^2,
\enea
and
\beq
M_D=2.3\times 10^{10} M_{\odot}\frac{[M_{vir}/(3\times
10^{11}M_{\odot})]^{3.1}}{1+[M_{vir}/(3 \times 10^{11}M_{\odot})]^{2.2}}. 
\label{eq:1.5}
\eneq
By means of equations  (\ref{eq:0.3})-(\ref{eq:1.5})  we can derive,  the  URCH halo velocity  $V_h^{URC}(r/R_D, L)$.\\
The  first  step  is to  investigate  the  $V_h(R_{1/2})$ - $R_{1/2}$ relationship, that  holds also  for objects of different Hubble types.  In Fig. (\ref{fig:1}) we  plot the URCH   as a red line and the relation for Sample A   as black  points. It is clear that  both individual and URC halo velocities  do correlate with  $R_{1/2}$,   well matching  the  W+10 (M+07) relation (shown as black line). \\
The further step is to investigate  the  full profile $V_h(r)$   at ``intermediate radii" $ \max\left[\frac{1}{3}R_D, 1 \mbox{kpc}\right]\leqslant \,r \, \leqslant 6 R_D$.  Each of the 116  spirals contributes with  2-4 independent kinematical  measurements. In Fig. (\ref{fig:2}) we plot $V_h(r)$   obtained from the RCs of Sample A,  once  that the luminous component has been removed as explained above.
We also plot, as two red  lines,  the URCH velocity profiles  of  spirals with luminosities similar to those of the objects of  Sample A  \footnote{The related virial masses  are  $M_{vir}/M_{\odot}=1.17 \times 10^{11}$, $7.7 \times 10^{11}$.}, and to the great majority of the 60  spirals of  the   W+10 sample. Notice that, given the Schechter-like form of the luminosity function of spirals, in the W+10 sample there are  only an handful of big galaxies. \\
In the low luminosity range we plot  $V_h(r)$ for the individual RCs  of Sample A (black points) and for the corresponding URCH profiles, and compare them with the W+10 relationship.
Let us recall that we did not investigate the  {\it individual}\,  $V_h(r)$ of  luminous spirals since their  dark-luminous  RC decomposition  is quite  uncertain. However, we have investigated  the  high luminosity objects  by  means of the URCH, that we show in Fig. (\ref{fig:4})  alongside  with  the  W+10 relationship.
Summarizing, for    spirals in halos  with $M_{vir}> 1.9  \times 10^{12} M_\odot$, we derive $V_{h}(r) $   by means of the URCH,  for  halos of  lower  mass, instead, we derive  $V_{h}(r)$  from  1) the  URCH (red lines) and 2) the mass modelling of the  individual RCs  of Sample A (black points). In both cases,  the (derived) halo velocities are  compared with the W+10 relationship.

\section {Universal halo  velocity profile and implications} 

\begin{figure*}[htbp]
\includegraphics[width=14cm,height=16cm]{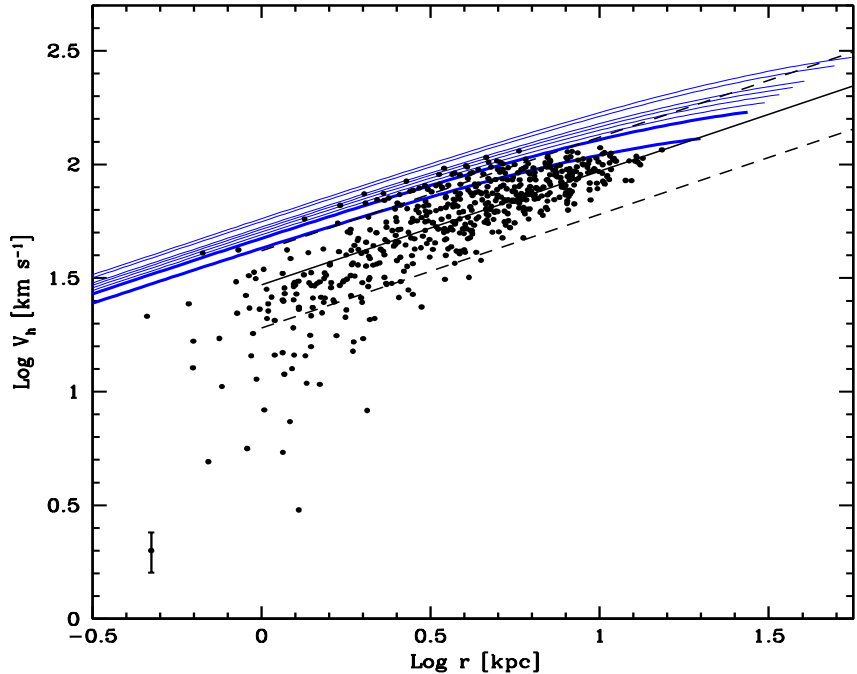}
\vskip -3.5cm
\caption{The W+10 halo velocity-radius  relationship (black line)  with its 1-$\sigma$ scatter (black dashed lines) plotted with  1) the halo velocities of  our Sample A of 116 spirals (black points), 2) the NFW halo velocity profiles (blue lines,  with the  thick  curves corresponding to objects  with  mass comparable with those of the  individual Sample).}
\label{fig:3}
\end{figure*}

We find for Sample A and also  from the  URC, that: $V_h(R_{1/2})\propto R_{1/2}^{0.5}$,    very similar to the W+10 relation. This  result is not a new one, it  arises as effect  of the  particular   dependencies of  the dark and luminous structural parameters  on  luminosity and  disk length scale.   We can show this by  looking at the Radial Tully Fisher relation (\cite{yego}).  From their eq. (9) one has $L_I \propto V(R_{1/2})^{2.55}$,  where $L_I$ is the I-band luminosity,  considering that for spirals we have: $R_{1/2}\propto R_D\propto L^{0.5}$, and  that from the mass modelling of individual objects  $V(R_{1/2})/V_h(R_{1/2}) \propto  V_h(R_{1/2})^{0.2}$ (see PS90, \cite{persal90b,sal}, i.e. smaller galaxies have a larger fraction of DM),   one arrives  to  $V_h(R_{1/2})\propto R_{1/2}^{k}$, with $k$ very near to the value of $+0.5$,  claimed by W+10  (see Fig. (\ref{fig:1})).  \\
However, with the  complete  data from the URC  shown in Fig. (\ref{fig:3}) we realize  that  $V_{h}(r)$  in Spirals of different  luminosities, has  not an unique  profile,  neither always it is a  power law   with exponent 1/2.  However, at  low luminosities,  the halo velocity individual  contributions and the URCH profiles are in good agreement with the  W+10  result  (see also  M+07). \\  
At highest luminosities (see Fig. (\ref{fig:3})) the W+10 relationship cannot be reproduced  by the  URCH profiles. However let us stress that in these cases  $V_h(r)$  depend  significantly (especially in the  W+10 work) on the  assumptions adopted  in the mass modelling. Moreover, the URCH profiles are obtained from  hundreds of RCs, while $V_h(r)$  in W+10 only from  few (see W+10).  \\
As a result, in a first approximation we can claim that  the W+10 relationship, in the radial range 1 kpc $\leq  r \leq 30$ kpc,  is a projection, on the ($r$, $V_h(r)$) axes of a more complex $V_{URC} (r/R_{opt}, L)$ relationship   in which the structural quantities   $V_{URC} (1, L)$, $L$ and $R_D$ are related in a specific way. This suggests the existence of a number of scaling laws between disk scalenghts, disk mass, halo structural parameters and galaxy luminosity,  which by the way have been also observed through the analysis of accurate mass models in large samples of galaxies (e.g. S+07). \\
The W+10 relation URC-supported, becomes  very useful to test  whether  DM halos in spirals are consistent with NFW density  profile.
In this case let us remind that  (see \cite{nfw}):
\beq
V^2_{NFW}(r)=V^2_{vir}\frac{g(c)}{xg(cx)},
\eneq 
where $x=r/R_{vir}$ is the radial coordinate, $g(c)=\left[ln(1+c)-c/(1+c) \right]^{-1}$, and for the concentration parameter $c$ we take
\beq
c(M_{vir})=9.2\left(\frac{M_{vir}}{10^{12}h^{-1}M_{\odot}} \right)^{-0.09},
\eneq
(see \cite{dmaw,klypin}). The halo mass inside a radius $r$ is given by: 
\beq
M_{NFW}(x)=M_{vir}\frac{g(c)}{g(cx)}.
\eneq
In Fig. (\ref{fig:3})  we plot $V_{NFW}(r, M_{vir}) $ for  the  radial  and halo mass ranges identical with those investigated above (more specifically we consider 8 masses equally log-spaced in the range:  $7.7 \times 10^{11}\leqslant M_{vir}/M_{\odot}\leqslant 1.1\times 10^{13}$). 
The resulting  halo  velocity curves $V_{NFW}(r)$  occupies a  very well defined portion in the ($r$, $V_{h}(r)$) plane (blue lines).  However, this region is mostly off that   defined  by  1) the  W+10 relationship, 2) the similar  relationship derived in this paper by a larger number of  individual RCs (of low luminosity objects) and 3) a specific  projection of the  URCH (given in Fig. (\ref{fig:2})).
A disagreement of the NFW halo velocity profiles  with  the actual  kinematics of spirals is not  new (e.g. \cite{Gentile}). However, here, we make a step further by investigating  an unprecedented large number of low luminosity objects.  We derive  their $V_h(r)$  profiles from disk kinematics   by adopting  only weak and reasonable assumptions.  These profiles result incompatible with the  NFW  profiles {\it for any  value  of their masses and their  concentrations}.

\section{Discussion}

We have compared the W+10 $(V_h(R_{1/2})$ -$R_{1/2})$  relationship  (see also M+07), with the  URCH  and with the kinematics  of 116 spirals of low luminosity objects.  We find that the  W+10 ``unique'' velocity  profile is essentially in agreement with the URC paradigm of which it is a sort of projection.  The W+10 relation alongside  its not negligible  scatter of  0.25 dex, can be identified as the projection of the URC  on the  ($r$, $V_h(r)$) plane.  This  implies the existence of a number of  scaling laws between  disk scalenghts, disk mass, halo structural parameters  and galaxy luminosity,  observed also analyzing accurate mass models in large samples of galaxies. \\ 
We claim that the distribution of dark and luminous matter in galaxies is such that  at  any chosen  radius $r$, $V_h(r)$ takes, in different galaxies of  different mass,  almost   the  same value.  The simplest explanation is that the more massive is  the  dark  halo, the lesser dense and concentrated it turns out to be. The net effect  is that the dark  mass inside a chosen  physical radius  is weakly dependent of the total  halo mass.
The physical explanation of this still lags: we realize that the  dark and the  luminous matter are distributed in a related way, which is  not obvious,  in view of their very different nature.  
It is worth to point out that  the URCH profiles for  virial masses M$_{vir} > 3 \times 10^{12}M_\odot$,  are discrepant with the W+10 relationship,  also considering  its (quite large)  scatter (see Fig. (\ref{fig:4})). Although in  the W+10 sample there is an obvious shortage  of such big objects, it  would seem  that  for the  about   5\%  top  luminous objects, the halo  velocity  profile implied by W+10 does not agree with the URC mass model and it  indicates a  larger  amount of DM. This issue will  be clarified  only with a bigger sample of individual RCs. 
At low masses, low circular  velocities and low   radii (1 kpc $<$ r $<$ 30 kpc), we find the existence of an ``unique'' $V_h(r)$ profile whose physical meaning has to be understood. This is amazing in that   by  adopting  the  radial units,  i.e. kpc,  we lose any autosimilarity of the velocity/mass  profiles,  present when   the radial coordinate is measured in terms  of R$_{vir}$ or  R$_D$.
We can use the W+10 relationship as a Cosmology Test: in fact,  when plotted in  radial  physical units,  the  NFW  velocity profiles  lie in a very small region, almost independently of their halo masses and concentration. Such  a region, however, turns out to be clearly  outside  that indicated by  the derived  halo velocity profiles in  W+10, in the URCH and in our sample of individual RCs. 
Especially  at low luminosities, where the NFW velocities are 0.2-0.4 dex higher,  the observed discrepancy is  very severe in that, being the contribution of the luminous matter quite negligible,  $V_h(r)$ is virtually derived in a model independent way, and it  almost coincides  with the observed $V(r)$.
Of course, the inconsistency  between RCs and NFW halo + disk mass model is already  well known, even if  it is  proved only  in  a  quite limited number of test cases (\cite{salfri}).  Here we have opened  the possibility to investigate such a crucial  issue  in  direct  way and by means of a large number of  objects. It is worth recalling  that also within  $\Lambda$CDM scenario there are several ways to modify the original NFW halo profiles to be compatible with observations, see \cite{maccio, governato, tonini} and references therein.

\end{document}